\documentstyle[11pt,paspconf]{article}

\begin{document} 
 
\title{Primordial Cosmological Fluctuations on Galactic Scale}  
\author{Claudio Firmani, Xavier Hern\'{a}ndez, and Vladimir  
Avila-Reese}  

\affil{Centro de Instrumentos, U.N.A.M., \\
Apdo. Postal 70-186, 04510 M\'exico D.F., Mexico }

\affil{Instituto de Astronom\'{\i}a, U.N.A.M., \\
Apdo. Postal 70-264, 04510 M\'exico D.F., Mexico}
 
\affil {Presented at the Workshop "Observational Cosmology: From Galaxies to Galaxy Systems", July 1995. Accepted in ApLC (November 1995).}
  
\begin{abstract}  
The galactic evolutionary theory is now sufficiently mature to provide  
information about the galaxy formation phase. From evolutionary models we  
establish a link between the present features of late type galaxies and the  
protogalactic density fluctuations. This link is used to estimate the  
fluctuation power spectrum amplitude at galactic scales. The agreement with  
the power spectrum derived from galaxy distribution and extrapolated 
to galactic scales, is satisfactory. Meanwhile, the comparison with respect to the standard cold dark matter power spectrum normalized to COBE satellite meassurements reveals only a marginal agreement. Among the most interesting results, the  Tully-Fisher relation and the constant central brightness pointed out by Freeman  (1970) appear to be the natural consequences of the initial cosmological conditions.  
\end{abstract}
  

\section{Introduction}  
  
Modern cosmology offers the key concepts for understanding how, from an  
almost homogeneous sea of particles and radiation, the universe developed  
structures as complex as the galaxies, clusters and super clusters of  
galaxies that we observe today. At this point, it appears logical to  
question ourselves about the natural initial conditions that cosmology could  
provide for the astrophysical processes of galaxy formation and evolution.  
This question would result interesting only if the present structure of  
galaxies retained some information related to the cosmological physical  
conditions prevalent at the epochs when galactic seeds appeared; that is, if  
strong dissipative processes did not totally erase the initial cosmological  
conditions. If this were the case, the following question immediately  
arises: Is it possible to establish a link between cosmological models and  
the properties of galaxies seen today? It is the aim of this work to find  
precisely such a link, using for this purpose a galactic evolutionary theory.  
  
The mayor ingredient relevant to the problem of structure and galaxy  
formation is the prediction from cosmology of an initial fluctuation type  
and distribution. The most popular cosmological approach, due to its  
simplicity and predictive capability, is the inflationary Cold Dark Matter  
model $\left( CDM\right) $ including their variants, and the usual assumption is that the  
primordial fluctuations have a Gaussian distribution (is it really a solid 
prediction of the inflationary models?), with a scale invariant density  
contrast.  
  
Once a Gaussian nature is assumed for the density fluctuation field, its  
statistical properties can be determined in terms only of the fluctuation  
power spectrum $\left( PS\right) $. It has become customary to compare different cosmological scenarios at the level of their respective power  
spectra, linearly extrapolated to $z=0$. In this way, at some epoch, say the  
recombination time, one has the initial conditions for structure and galaxy  
formation.  
  
The normalization of this $PS$ is not provided by the theory, and has been  
estimated on intermediate scales by the galaxy distribution and, more  
recently, at very large scales, by the temperature fluctuations of the  
cosmic microwave background radiation $\left( CMB\right) $. The link we shall  
establish between cosmological models and the properties of galaxies, indeed  
offers a method to find the amplitude of the initial density fluctuations  
at the smallest cosmological relevant regions ---the galactic scales. In this  
way, the different $PS$ predicted by distinct cosmological models and  
normalized to the $COBE\ CMB$ measurements, can be tested using galactic  
evolution, at the smallest scales, where maximal variations among them  
typically occur.  
  
\section{Galactic Evolution and the Epoch of Galaxy Formation}  
  
Our goal is to find the typical formation redshift $z_f$ of galaxies with a  
given mass $M$, using for this the galactic evolutionary theory.  
Here we treat galactic evolution within the framework introduced in Firmani  
et al. (1995). The global evolutionary models presented there are based on  
the local ones developed by Firmani \& Tutukov (1992, 1994).  
  
The initial conditions from which the global models start include a  
spherical distribution of $DM$, and $depending$ $on$ $this$, they predict the  
properties of a late type galaxy as seen today. We pay special attention to  
the Tully-Fisher relation and the constant central surface brightness  
pointed out by Freeman (1970), because they seem to be more directly related  
to the cosmological conditions.  
  
It is very difficult to obtain an explicit solution to the problem of non-linear evolution of the primordial density  
fluctuations and the resulting $DM$ halo structures  
(see for example Flores \&\ Primack (1994), Moore (1994), Klypin et al. (1995)).  
In order to avoid complicated dynamical calculations which in any case  
involve statistical uncertain primordial density fluctuation profiles, in the frame of our simplified approach, we assume  
that, during the non-linear evolution of the density fluctuations, a violent  
relaxation virializes the $DM$ halo to a truncated isothermal sphere with a  
core (Binney \& Tremaine (1987), Bachall \& Soneira (1980) ). We also  
experimented with alternative halo profiles (e.g. Katz (1991)). The density  
profile parameters of the virialized halo can be easily related through  
energy conservation to the radius of the primordial fluctuation at the  
maximum expansion, if a $top-hat$ configuration is assumed for this phase.  
This relation reduces the degree of freedom of the virialized halo  
configuration to one, which we fix in order to obtain a flat rotation curve  
at the end of the evolution. This simple procedure allows us to construct a  
virialized halo starting from $M$ and $z_f$.  
  
One assumes that after the virialization of the $DM$ halo, all the baryonic  
gas trapped in it, cools and falls to the center forming a centrifugally  
supported disk where the stellar populations form, evolve, and interact with  
the interstellar medium. Here the disk mass fraction, in principle, may be  
assumed equal to the fraction of baryonic matter $\Omega _b$ which is well  
predicted by big bang nucleosynthesis (BBN). The initial angular momentum,  
measured by the dimensionless angular momentum $\lambda $, is taken from the  
tidal torque theory and numerical simulations (Peebles 1969, Fall \&\  
Efstathiou (1980), Efstathiou \&\ Barnes (1987)), having an average value of $%
\lambda \approx 0.05$. We do not consider the possibility of mayor mergers  
after the galactic disk is formed. Since late type galaxies are typically  
field objects, or located at cluster peripheries, we think this last  
assumption is quite reasonable.  
  
Now, given a total mass $M$ and following the above stated  
assumptions of a final flat rotation curve and a baryon content close to the  
BBN prescription, one selects those models which satisfy, through the galactic  
evolution, the Tully-Fisher relation (normalized to the Galaxy) and the  
Freeman law. This procedure fixes $z_f$ and $\lambda $ completely.  
  
The results, obtained using an isothermal halo, $h=0.5\ $ and $\Omega  
_b=0.05 $, are plotted in Fig.1 (continuous line). As shown in Fig.1, the  
mass roughly scales with $z_f$ as $M\propto \left( 1+z_f\right) ^{-6}$ at galactic scales. On other hand, it is possible to estimate such a relation for the hierarchical models, using for this the Press-Schechter formalism (Press \&\ Schechter (1974)). In Fig. 1 we also show the average $z_f$ of collapsed objects with masses $M$ for a $COBE$-normalized standard $CDM$ power spectrum.  
  
The rather good agreement found between our predictions of galaxy formation  
epochs and the Press-Schechter formalism applied to a standard $CDM$  
cosmology is quite encouraging of the validity of our approach. We feel it  
is now safe to state that the Tully-Fisher scale relationship and the  
constant central disk brightness found by Freeman are the fossil remains of  
the physical conditions prevalent during the epoch of galaxy formation. That the slope of the Tully-Fisher relation is related to the cosmological conditions was previously suggested (e.g. Faber (1981)), but the proportionality coefficient was not considered.  
  
It is encouraging that we obtain the same value for $\lambda $ which the  
numerical simulations predict. Inverting the procedure, if we fix $\lambda  
=0.05$, we can predict a disk mass fraction $\approx $ $0.05$, which results  
close to the value predicted by the BBN.  
  
Several experiments with alternative halo profiles have been taken into  
consideration, with poor results in terms of final rotation curves.

\begin{figure}
\vspace{3.0cm}  
\caption{Formation redshift vs. mass from disk galaxy evolutionary models (solid line), and from the standard $CDM$ model (dashed line).}  
\end{figure}

\section{The Normalization of Power Spectrum at Galactic Scales}
  
From galactic evolution we were able to determine the formation redshifts of  
galaxies. Now, using the Press-Schechter formalism in reverse, and the  
spherical collapse model, it is straightforward to find the amplitude of the  
initial density fluctuations at galactic scales.  

\begin{figure}
\vspace{4.4cm}  
\caption{Calculated power spectrum in  the galactic region, thick line, compared with the figure of Stompor et al. (1995), which shows various spectrum estimates from galactic distribution observations, as well as several theoretical power spectra}
\end{figure}
  
In Fig.2 we plot the $PS$ obtained from galactic evolution at wave numbers  
between $1$ and $10~h~Mpc^{-1}$. Also plotted are several observational  
estimates on larger scales and theoretical $COBE$-normalized power spectra  
(from Stompor et al. 1995). It is clearly seen that our results are a  
natural small scale extension of the $PS$ estimated from galaxy distribution  
observations. Though it is still early to claim definitive results, we find  
our approach useful for constraining cosmological models at galactic scales.  
As was pointed out by Wambsganss et al. (1995),  since cosmological models  
have an assumed $PS$ that is normalized to pass through the region measured  
by the $COBE$ satellite at very large scales, and given that the slope of  
the $PS$ is a model-dependent feature, then the maximal variations among  
them naturally occur at the smallest scales (see Fig.2). The relevance of  
finding further estimates of the primordial fluctuation amplitude on scales  
as far removed as possible from the $COBE$ measurements, for example  
galactic scales, is at this point obvious.  
  
\section{Conclusions}  
  
We can summarize the results of this work as follows:  
  
1) The density fluctuation $PS$ inferred from galactic evolutionary models  
defines a sequence in the range $1<k<10~h~Mpc^{-1}$which shows itself as the  
natural extension of the estimates obtained from the observations of galaxy  
distribution.  
  
2) The Tully-Fisher relation (both, the slope and the proportionality coefficient) and the constant central brightness of late  
type galaxies appear to be consequences of the cosmological origin of  
galaxies. Both relations may be considered fossil information of the  
primordial  density fluctuation field.  
  
3) On the frame of a Gaussian distribution of the primordial density  
fluctuations, our results show in the fluctuation spectrum less power than  
the standard $CDM$ model, and lean slightly towards $CDM$ models with a  
non-vanishing cosmological constant.

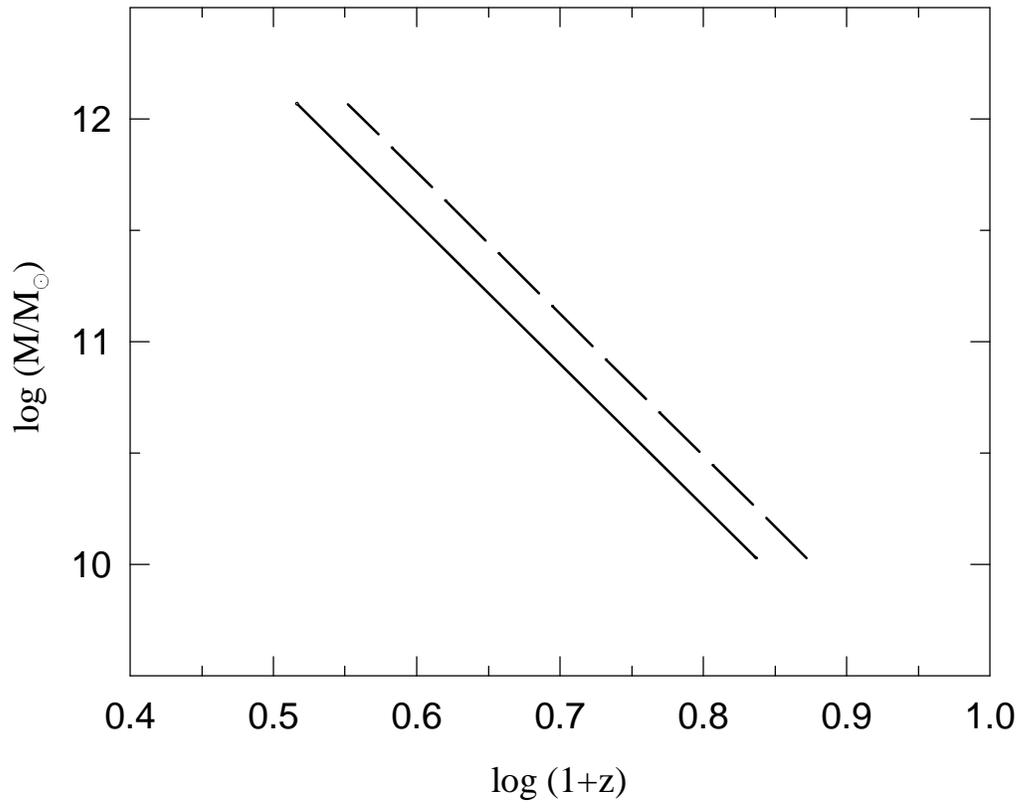
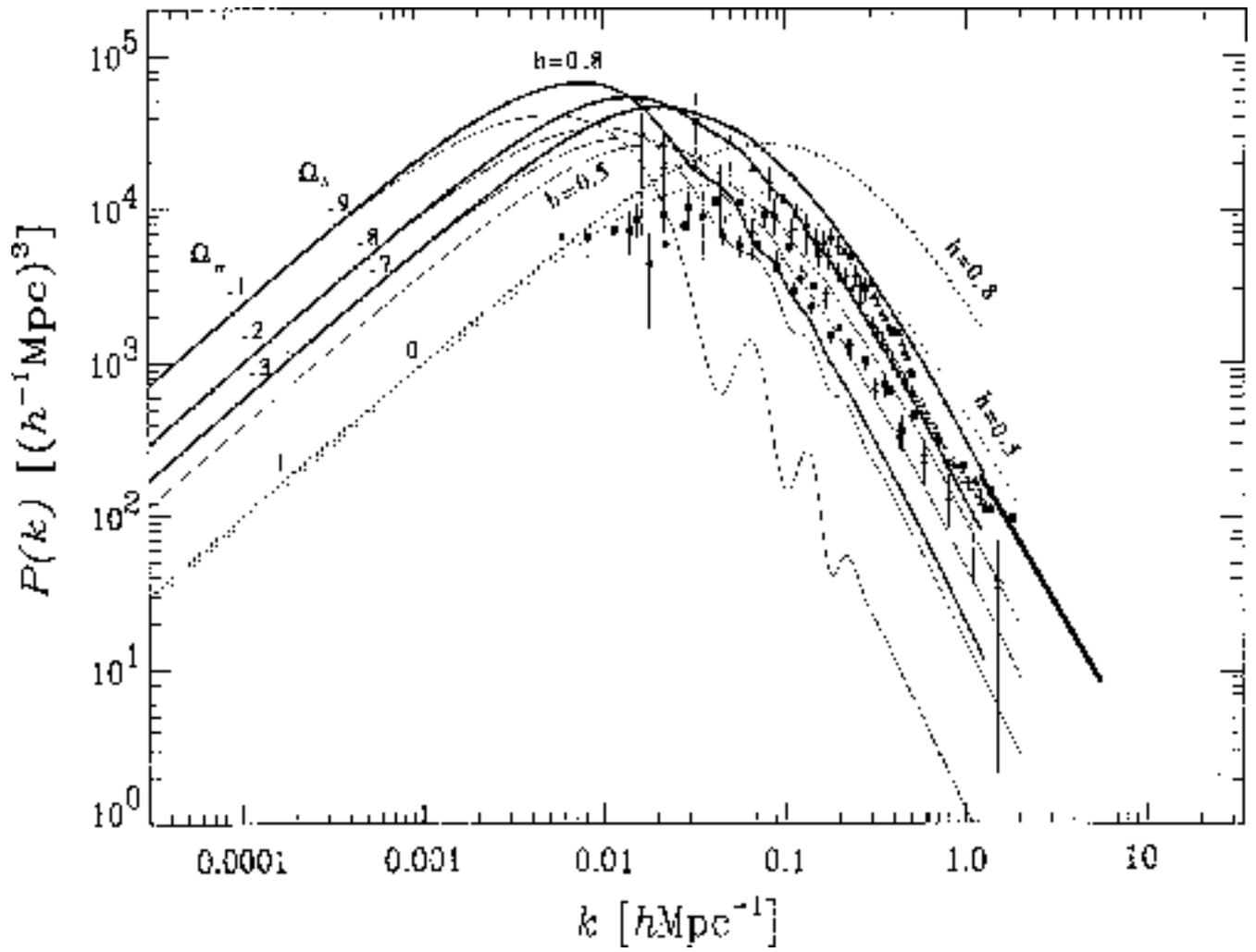
\end{document}